\documentclass[namedreferences]{SolarPhysics}
\usepackage[optionalrh,solaenum,natbib]{spr-sola-addons} 
\usepackage{graphicx}                    
\usepackage{color}                       
\usepackage{url}                         

\usepackage{solaheader}	
\newcommand{\solareceiveddate}{date} 
\newcommand{\solaaccepteddate}{date}
\newcommand{\solaonlinedate}{date}
\newcommand{\Xdoi}{10.1007/s11207-012-9985-9}	
\newcommand{\ion}[2]{#1\,{\sc #2}}

\begin{document}

\begin{article}

\begin{opening}

\title{Automation of the Filament Tracking in the Framework of the HELIO project}

%
\author{X.~\surname{Bonnin}$^{1}$\sep
        J.~\surname{Aboudarham}$^{1}$\sep
        N.~\surname{Fuller}$^{1}$\sep
        A.~\surname{Csillaghy}$^{2}$\sep
        R.~\surname{Bentley}$^{3}$\sep
       }

%
\runningauthor{X.Bonnin et al.}
\runningtitle{Automation of Filament Tracking}

%
  \institute{$^{1}$ LESIA, Observatoire de Paris, CNRS, UPMC, Universit\'e Paris-Diderot, 5 place Jules Janssen, 92195 Meudon, France
                     email: \url{xavier.bonnin@obspm.fr} email: \url{jean.aboudarham@obspm.fr} email: \url{nicolas.fuller@obspm.fr} \\ 
			 $^{2}$ Institute of 4D Technologies, FHNW, Steinackerstrass 5, CH-5210 Windisch, Switzerland
			 		 email: \url{andre.csillaghy@fhnw.ch} \\           
             $^{3}$ MSSL, University College London, Hombury St. Mary, Dorking, Surrey RH5 6NT, U.K.
			            email: \url{rdb@mssl.ucl.ac.uk}  \\
             }

\begin{abstract}
We present a new method to automatically track filaments over the solar disk. The filaments are first detected on Meudon Spectroheliograph H$\alpha$ images of the Sun, applying the technique developed by \citeauthor{FullerSoPh2005} (\textit{Solar phys.} \textbf{227}, 61, 2005). This technique combines cleaning processes, image segmentation based on region growing, and morphological parameter extraction, including the determination of filament skeletons. The coordinates of the skeleton pixels, given in a heliocentric system, are then converted to a more appropriate reference frame that follows the rotation of the Sun surface. In such a frame, a co-rotating filament is always located around the same position, and its skeletons (extracted from each image) are thus spatially close, forming a group of adjacent features. In a third step, the shape of each skeleton is compared with its neighbours using a curve-matching algorithm. This step will permit us to define the probability $[P]$ that two close filaments in the co-rotating frame are actually the same one observed on two different images. At the end, the pairs of features, for which the corresponding probability is greater than a threshold value, are associated using tracking identification indexes.On a representative sample of filaments, the good agreement between automated and manual tracking confirms the reliability of the technique to be applied on large data sets. Especially, this code is already used in the framework of the \textit{Heliophysics Integrated Observatory} (HELIO) to populate a catalogue dedicated to solar and heliospheric features (HFC). An extension of this method to others filament observations, and possibly the sunspots, faculae, and coronal holes tracking can be also envisaged.
\end{abstract}

%

\end{opening}

%
\section{Introduction}
     \label{S:Introduction} 

The heliophysics community embraces a number of existing disciplines -- solar and heliospheric physics, magnetospheric and ionospheric physics for the Earth and other planets -- and thus must deal with very large sets of heterogeneous data from ground- and space-based instruments. The \textit{Heliophysics Integrated Observatory} (HELIO: \url{http://www.helio-vo.eu}) is a virtual observatory dedicated to the solar physics and heliophysics \citep{Bentley_AdSpR_2011}. HELIO provides a distributed network of services, which helps the researchers to easily mine relevant information and data. Especially, the \textit{Heliospheric Feature Catalogue} (HFC: \url{http://voparis-helio.obspm.fr/hfc-gui/index.php}) is a database-oriented service that allows access to a large amount of solar and heliospheric features data. Extraction of feature information stored in the HFC is realized using an increasing number of recognition codes \citep{FullerSoPh2005,Zharkov_jasp_2005,Barra_aa_2009,Krista_solphy_2009,Lobzin_sw_2009,Higgins_aph_2011}.\\

We present here a new algorithm for the solar-filament tracking. It has been initially developed in the framework of the HELIO project in order to provide tracking data, as a supplementary information, to the description of filaments already available in the HFC. The filaments are large-scale structures of relatively dense and cool plasma suspended in the hot and thin corona. They are particularly visible on H$\alpha$ observations, where they appear as elongated dark features with several barbs on the solar chromosphere \citep{TandbergASSL1995,MackaySSR2010}. During their lifetime, their shape and intensity can be subject to a number of modifications, especially, part or all of a filament can sometimes undergo sudden disappearances (which may be followed in some cases by re-appearances) at some wavelengths, probably due to changes in their background characteristics -- temperature, pressure, \textit{etc.} -- or caused by eruptive process. Hence, this sudden and unpredictable behaviour can make filaments difficult to track over the Sun surface. However, following these features over time is relevant, notably for space weather, since erupting disappearances can play a role in the triggering of coronal mass ejections (CMEs) \citep{GilbertApj2000ApJ,GopalswamyApj2003}.\\
Many automated methods have been successfully developed during the last ten years to detect filaments, including \citet{GaoSoPh2002}, \citet{FullerSoPh2005}, \citet{BernasconiSoPh2005}, \citet{ZharkovaSoPh2005a}, and more recently \citet{JoshiSoPh2010}. In addition to their recognition algorithms, \citet{GaoSoPh2002,BernasconiSoPh2005,JoshiSoPh2010} also propose tracking capabilities that allow to identify disappearances by following feature locations day by day. In particular, \citet{GaoSoPh2002} have tracked filament disappearances over a full year, and the code of \citet{BernasconiSoPh2005} is currently applied to detect and track filaments on Big Bear Solar Observatory (BBSO) images, providing among others data to the \textit{Heliophysics Events Knowledgebase} (HEK: \url{http://lmsal.com/hek/}). In all cases the basics of the methods are quite similar: starting from heliocentric positions of the filaments detected at a given time (\textit{i.e.}, on a given image), they estimate the coordinates of their respective centroids on the previous and/or next adjacent images using equations that correct for the solar rotation \citep{Cox2000}. Then, they check if detected filament lies within a given circle area from the predicted locations; if it does, the two filaments are considered to be the same. In the case of \citet{BernasconiSoPh2005}, where no filament is found they then extend the search up to three days to confirm that the filament actually disappears or not. As highlighted by the authors, the 3-day search is motivated by the fact that filaments sometimes change shape so much between times of observation, that their location may fall out of the search area, temporarily loosing the tracking. A consequence is that during this period of time it is not possible to know if the filament actually disappears, or if its shape is just strongly deformed due to splitting or partial disappearance. A tracking algorithm based on time, position, but also shape matching would permit to distinguish between the different possible behaviour. This capability appears to be essential to detect, among others, erupting disappearance.\\

In our method, filaments are first detected on H$\alpha$ images by an automated recognition code that extracts, among other things, the pruned skeletons of filaments. A pruned skeleton can be defined as the thickest curve inside the feature area that preserves the shape topology; in terms of image processing it results of a thinning followed by a pruning operation \citep{GonzalezDip2002} applied on the feature pixels. Then, the coordinates of the resulting skeleton pixels are converted to a more appropriate reference frame that rotates with the solar surface. In such a frame, the skeleton of a filament observed on successive images appears as a cluster of close curves \citep{MouradianASPC1998}. Hence, a comparison in this frame of each skeleton curve with its closer neighbours using a curve-matching algorithm, allows us to calculate probabilities that all of these features actually belong to the filament. The algorithm demonstrates its ability to track filaments on large data sets, limiting the number of false tracking detections. Section \ref{S:Detection} will introduce the observations used to test the technique, and briefly describe the filament detection process; Section \ref{S:Tracking} is devoted to the explanation of the tracking algorithm; Section \ref{S:Performances} will present the resulting performances of the code. Finally applications will be discussed in Section \ref{S:Conclusions}.

\section{Detection of Filaments}
     \label{S:Detection}
     
\subsection{Observations} 
  \label{S:Observations}     
     
\begin{figure} 
 \centerline{a.\includegraphics[width=0.45\textwidth]{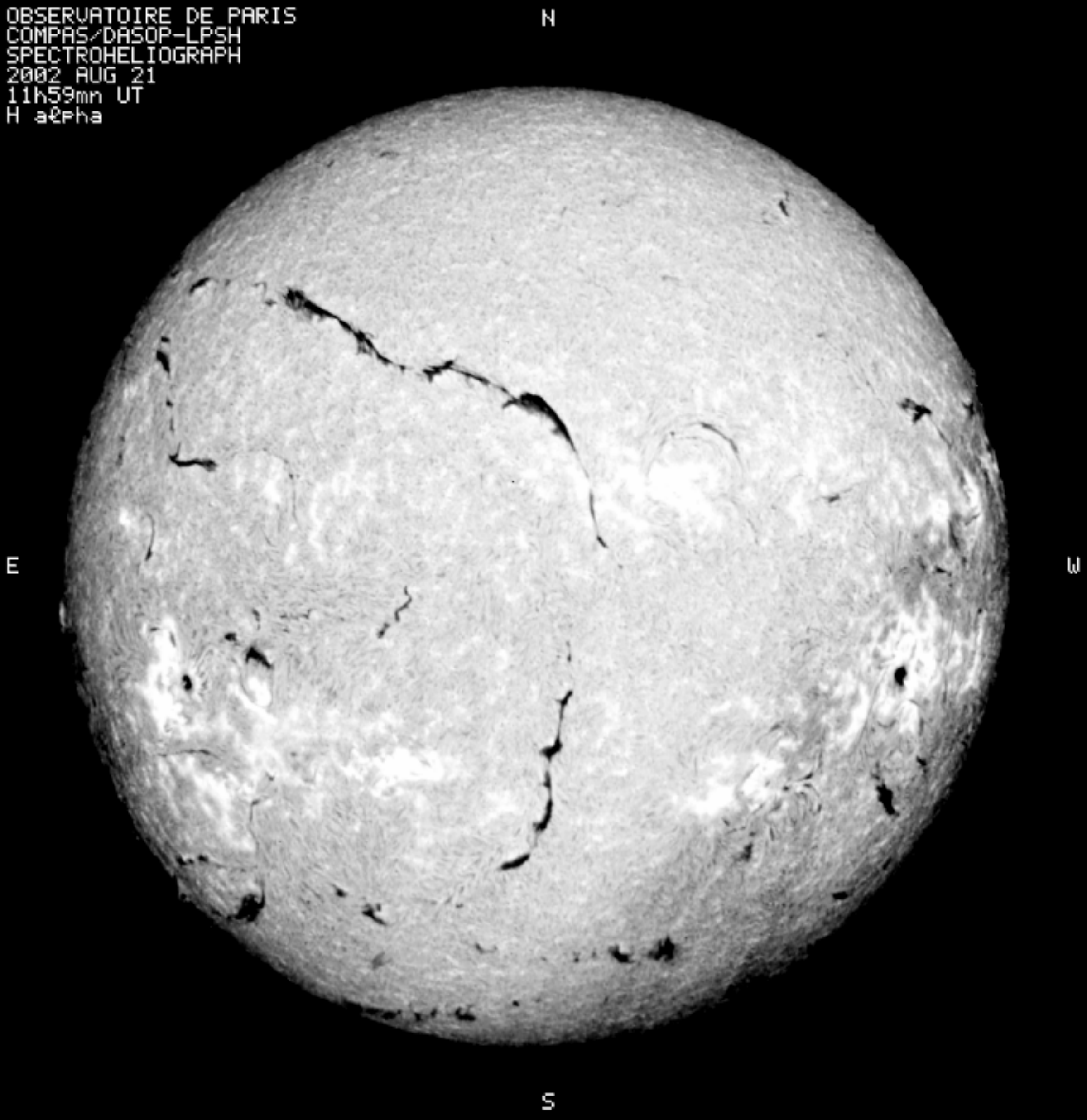} b.\includegraphics[width=0.45\textwidth]{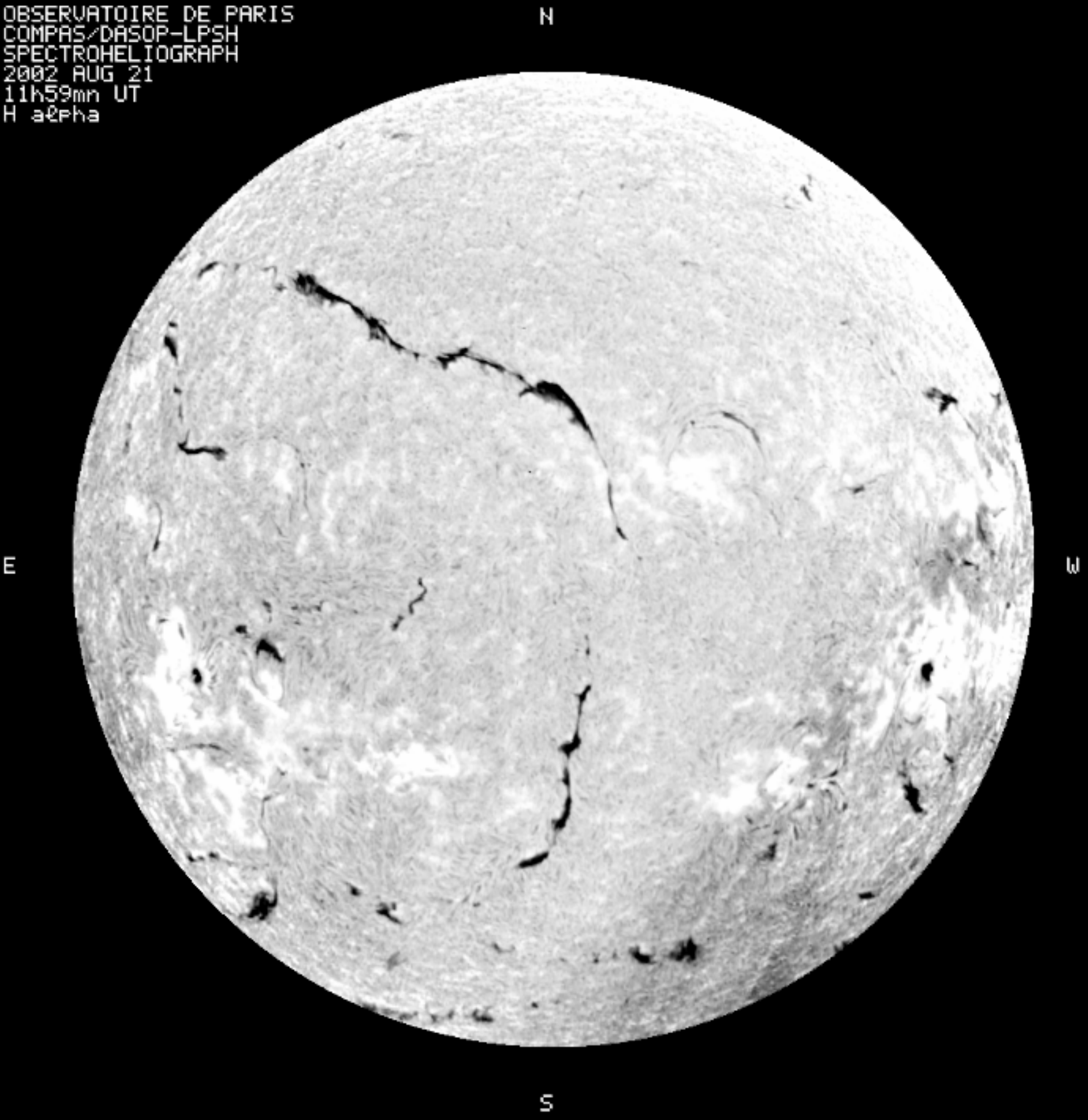}}
  \caption{a. Gray-scale image of the Sun observed in H$\alpha$ by the Meudon spectroheliograph on 21 August 2002 at 11:59:00 UT. The image is centred on the solar disk, and modified to have $x$- and $y$-axis respectively aligned with the Equator and rotation axis. Due to their cooler temperature, the filaments are seen like dark features on the solar surface. Centre-of-limb effects as well as dark faint lines, caused by dust particles on the optics, are also visible. b. Same image after cleaning processes.}\label{fig:MeudonSpectro}
\end{figure}

The Meudon spectroheliograph of the Observatoire de Paris performs daily observations of the solar photosphere and chromosphere at three wavelengths: blue wing of the \ion{Ca}{ii} K 393.4 nm line, or K$_{1v}$ ; centre of the \ion{Ca}{ii} K 393.4 nm line, or K${_3}$ ; and H$\alpha$ 656.3 nm. More than fifteen years of data are accessible from the BASS2000 website (\url{http://bass2000.obspm.fr/home.php}), which permits us to apply the tracking algorithm over a large period of time.\\
The images produced are 2D heliocentric projections of the full solar disk as seen from the Earth, with a typical size of 1024 by 1024 pixels, and a spatial resolution of $\approx$ 2.28 arcsecs. The origin of the reference frame corresponds to the disk centre, $x$-axis is aligned with the Sun's Equator and points towards the west limb, and $y$-axis is aligned with the rotation axis and points towards the North. Figure \ref{fig:MeudonSpectro} shows such an image before (left panel) and after (right panel) cleaning processes \citep{FullerSoPh2005} ; these pre-processes are required to avoid false detections, but also to optimize the efficiency of the region-growing process used by the detection algorithm, by increasing the contrast between the filaments and quiet-sun intensities. 

\subsection{Recognition Code} 
  \label{S:Recognition}
  
In order to retrieve the location and morphology of filament skeletons required to perform tracking, an automated recognition code is first run over the full data set available. We use here an algorithm developed in the framework of the \textit{European Grid of Solar Observations} (EGSO: \url{http://www.egso.org}) project, and successfully applied on H$\alpha$ data from the Meudon Spectroheliograph and BBSO \citep{FullerSoPh2005}. This code is now also used in HELIO to provide to the HFC, a description of the filaments detected on Meudon images. (Filaments detected on BBSO observations are planned to be added in HFC in few months.)\\
The recognition method requires three main steps to be completed:
\begin{enumerate}
\item Since images are sometimes blurred, which can reduce the efficiency of the filament segmentation, a Laplacian spatial filter \citep{RussIph2002} is first used to enhance the clearness of filament contours. 
\item Then, a region-growing algorithm \citep{GonzalezDip2002} permits us to group pixels of a same filament together. Starting from a seed region (found using an appropriate threshold value), the procedure searches the connected pixels for which the intensities are within a range defined by the mean and standard deviation of their neighbours.
\item A morphological-closing operator \citep{GonzalezDip2002} will finally merge resulting nearby regions that could be considered as a single filament on the segmented images.
\end{enumerate}

\begin{figure} 
 	\centerline{a.\includegraphics[width=0.8\textwidth]{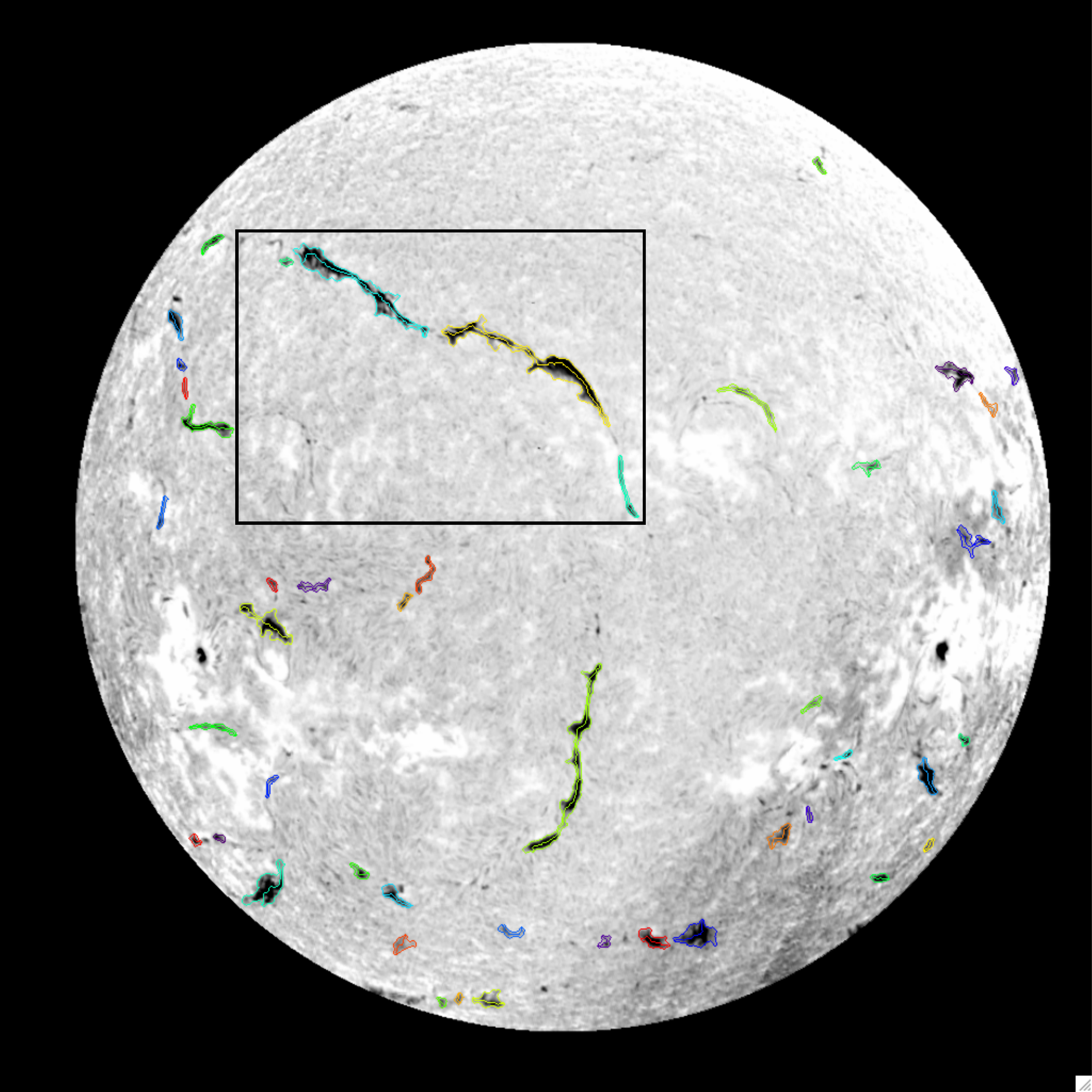}}
 	\centerline{b.\includegraphics[width=0.8\textwidth]{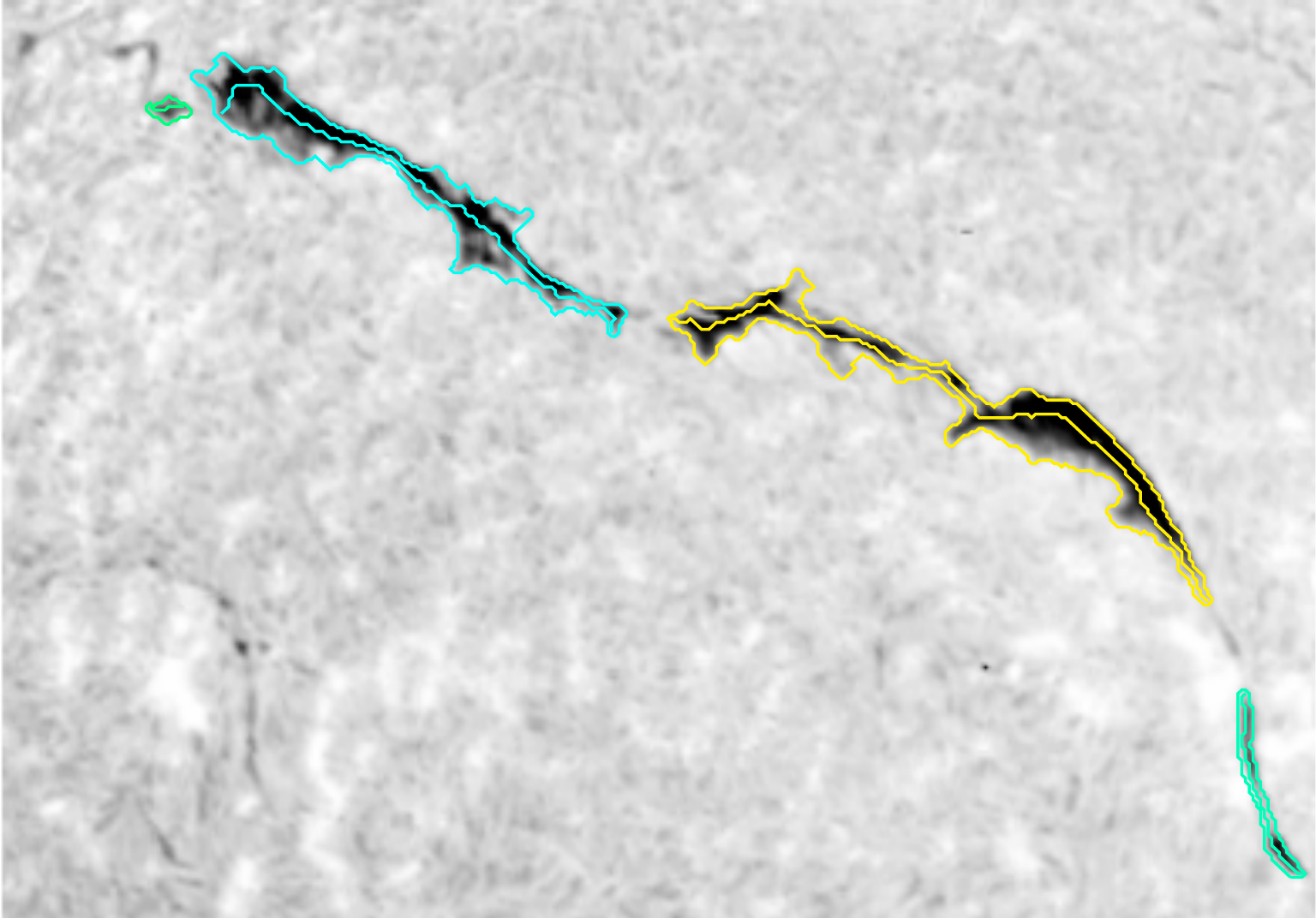}}
	\caption{a. Same image as in Figure \ref{fig:MeudonSpectro}b, but over-plotted with coloured contours and skeletons extracted from the recognition code. (The colors used here are randomly distributed to distinguish individual features detected.) b. Zoom on the area marked by the black box on the image displayed in Panel a.}\label{fig:filament}
\end{figure}

From the boundary detection, several parameters can be extracted concerning the intensity, location, or morphology of filaments. In particular, the algorithm uses thinning and pruning methods \citep{GonzalezDip2002} in order to compute skeleton shapes (results of a skeleton computation is illustrated on Figure \ref{fig:filament}). Two conditions are required to produce such shapes: all skeleton parts should be connected, and the full extent of the region must be contained in the skeletonized representation. Knowing this characteristic will permit us to define, among others, the length, the centre, and the curvature of the filaments. At the end, all of these parameters will be ingested in the HFC from where they can be downloaded using dedicated query interfaces.\\
For our purpose, the shape of the skeletons will serve as a matching criterion to identify a filament from an image to the following one.

\section{Tracking of Filaments}
     \label{S:Tracking}
     
\subsection{Description of the Algorithm} 
  \label{S:Algorithm}     

In this section we present in detail the tracking algorithm. In addition to the coordinates of skeleton pixels provided by the recognition code, two types of indices are also required to follow co-rotating features over time. The first index, called feature identification number $[\nu_{i}]$, is allocated to each filament $[i]$ extracted by the recognition code. All of its values are unique (the values of the feature identification number follow typically the order of detection of the filaments) in order to identify and retrieve in the HFC, a feature detected at a given time on a given image. The second index, called tracking identification number $[\tau_{i}]$, is assigned to each filament $[i]$ by the tracking code. As opposed to $\nu_{i}$ several filaments can have the same value for $\tau_{i}$, which will be an indication that these filaments are actually components of the same co-rotating feature observed on successive images. The values of $\tau_{i}$ are set by the tracking code, comparing and associating filaments two by two (as explained in Section \ref{S:Probability}). At the beginning of the execution, all of the tracking indexes $[\tau_{i}]$ are initially equal to the feature ones $[\nu_{i}]$ by default.\\

The main steps of the tracking method can be summarized as follows:

\begin{enumerate}
\item We select the time $[t_{\iota}]$ of the image $[\iota]$ for which we want to perform tracking. 
\item We then define a time window $[t_{0},t_{1}]$ centred on $t_{\iota}$ and spanning one Carrington rotation of the Sun, such that $t_{0} = t_{\iota} - 0.5T_{\mathrm{carr}}$ and $t_{1} = t_{\iota} + 0.5T_{\mathrm{carr}}$, where $T_{\mathrm{carr}} = 27.2753$ days. The choice of this time range makes sure that all of the filaments detected at $t_{\iota}$ will be fully tracked over a period of $0.5T_{\mathrm{carr}}$ days (which is actually the average time required to cross the solar disk).
\item We download the HFC information about filaments detected between $t_{0}$ and $t_{1}$; parameters required are typically the pixel coordinates of the pruned skeletons, but also their corresponding feature identification numbers $[\nu]$. (In practice, the corresponding tracking identification numbers  $[\tau]$ are also loaded to ensure a continuity with possible previous tracking runs.)
\item Since the pixel coordinates are given in the reference frame of images (see Section \ref{S:Observations}), we convert them to a more appropriate coordinates system as explained in the following section.
\item We apply a curve-matching algorithm in the new frame. The purpose of this algorithm is to compare the shape of all of the possible pairs of skeletons: if two skeletons have similar shapes, then they are assumed to be two distinct parts of the same filament moving over the Sun's surface.
\item When the comparison succeeds, we associate the two corresponding skeletons by allocating them the same tracking identification number. 
\item The same process is then repeated on the next image $[\iota+1]$ taken at time $t_{\iota+1}$, saving all of the tracking numbers $[\tau]$ defined during previous runs on the image $[\iota]$. This operation allows us to conserve tracking information from a processed image to the following.
\end{enumerate}

\subsubsection{Reference Frame used for Tracking} 
  \label{S:CarringtonMap}

The tracking for filaments detected at time $t=t_{\iota}$ is performed on a specific reference frame $\Re$ that follows the solar rotation between $t_0$ and $t_1$. Such a frame has the advantage to: i) stabilize the coordinates of co-rotating filaments around the same position, by correcting the translation of heliocentric longitudes for the rotation, ii) make appear all the segments of a filament visible between $t_0$ and $t_1$, which may suffer full or partial disappearance during its lifetime, iii) since along the first dimension time and space are intermingled, offer fast computation time by working in two dimensions instead of three. (The computation for all of the filaments detected over one solar rotation takes less than $\approx 30$ seconds on average on a $3.06$ GHz Intel duo core machine.)\\
For each image processed the detection code returns the heliocentric coordinates $(X,Y)$ of skeleton pixels (as defined in Section \ref{S:Observations}), which must be converted in the proper coordinates system associated to $\Re$ before being usable by the matching algorithm. To achieve this goal, the heliographic longitudes and latitudes $(\varphi_{h},\lambda_{h})$ of skeleton pixels on the Sun's surface are first computed using an appropriate transformation matrix \citep{ThompsonAA2006}, then the coordinates $(\varphi,\lambda)$ in the co-rotating frame $\Re$ are deduced using the relations: 

\begin{eqnarray}
\lambda = \lambda_{h}, \label{eq:coord1}\\
\varphi = \Omega_{\mathrm{carr}} (t_{0} - t) + (\varphi_{t_{0}} - \varphi_{h}) + \varphi_{DRC}(\lambda_{h}), \label{eq:coord2}
\end{eqnarray}

where $\Omega_{\mathrm{carr}}$ is the Carrington rotation speed in $degrees.day^{-1}$, $t$ is the time when the filament was detected in decimal days (\textit{i.e.}, the time of the observation), $t_{0}$ is the start time of the tracking period in decimal days, $\varphi_{t_{0}} = 360^\circ$ is the longitude of the central meridian in $\Re$ at $t_{0}$, and $\varphi_{DRC}(\lambda_{h})$ is a term that corrects the effects caused by differential rotation \citep{UlrichSoPh2006}. These effects are not significant here, but will become important when the algorithm will be run to track filaments from one solar rotation to the following (see Section \ref{S:SeveralRotations}). We note that the coordinates system defined here is quite similar to the Carrington one, but starting at $\varphi_{t_{0}}$.\\ 

\begin{figure} 
 \centerline{a.\includegraphics[width=0.50\textwidth]{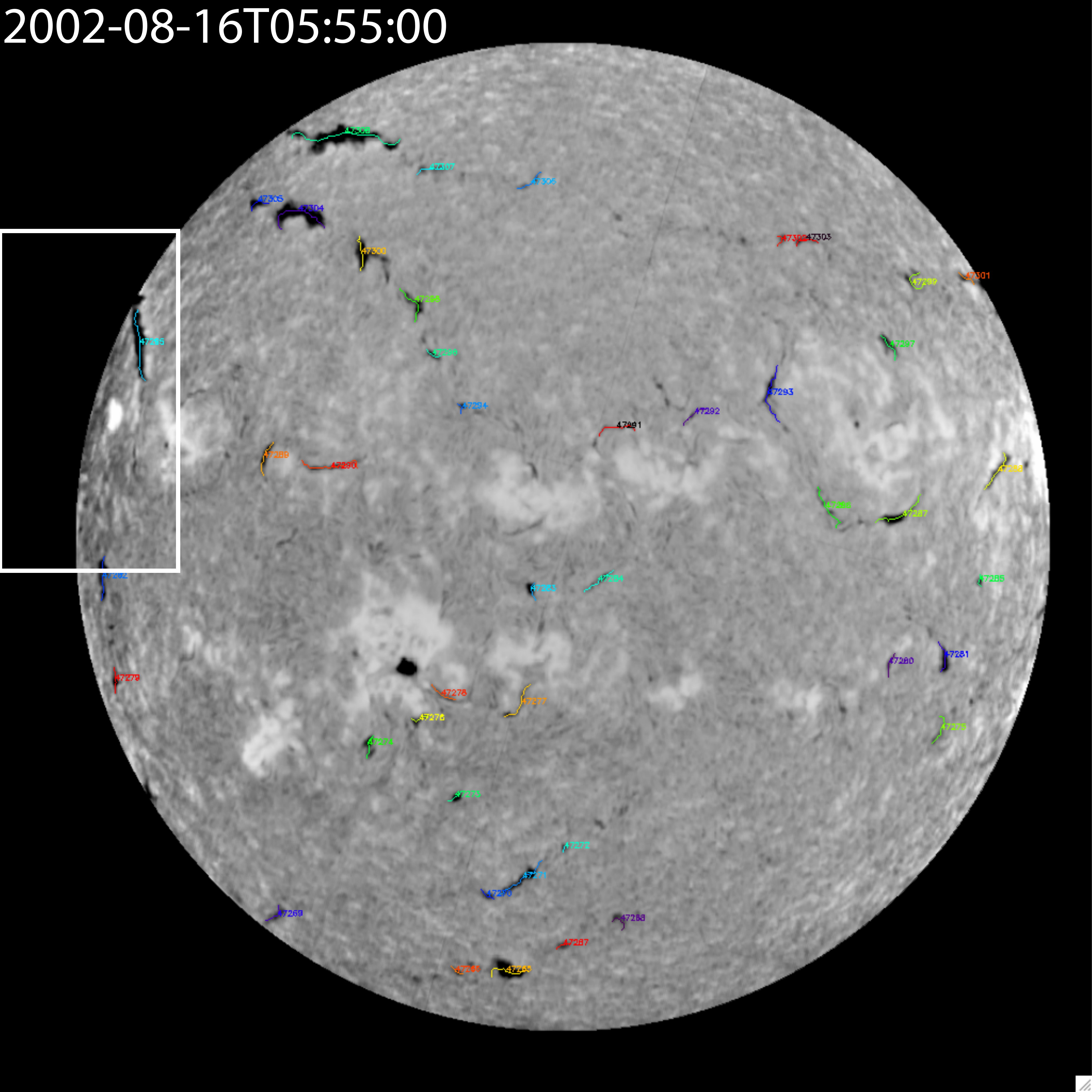} b.\includegraphics[width=0.50\textwidth]{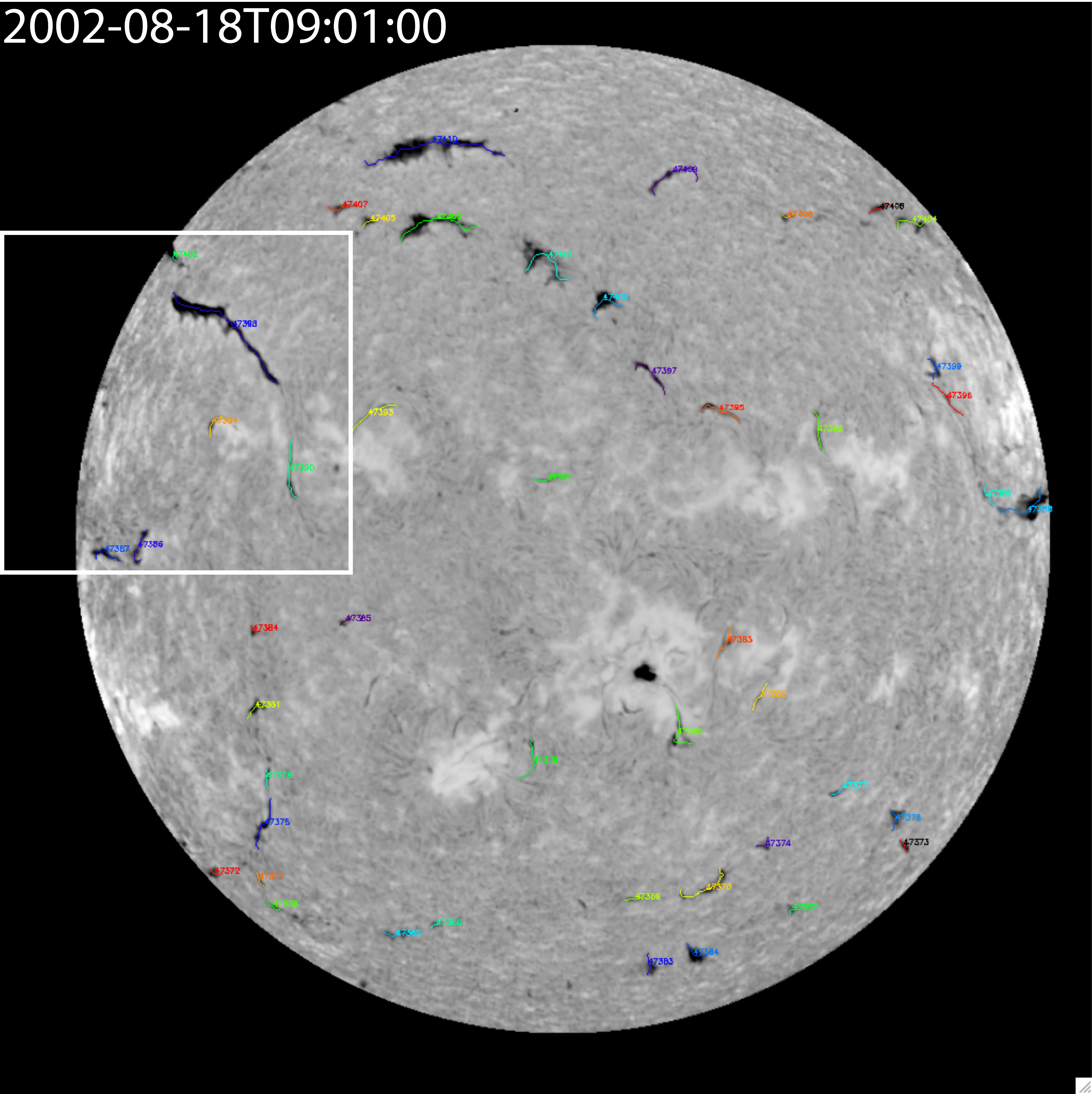}}
 \centerline{c.\includegraphics[width=0.50\textwidth]{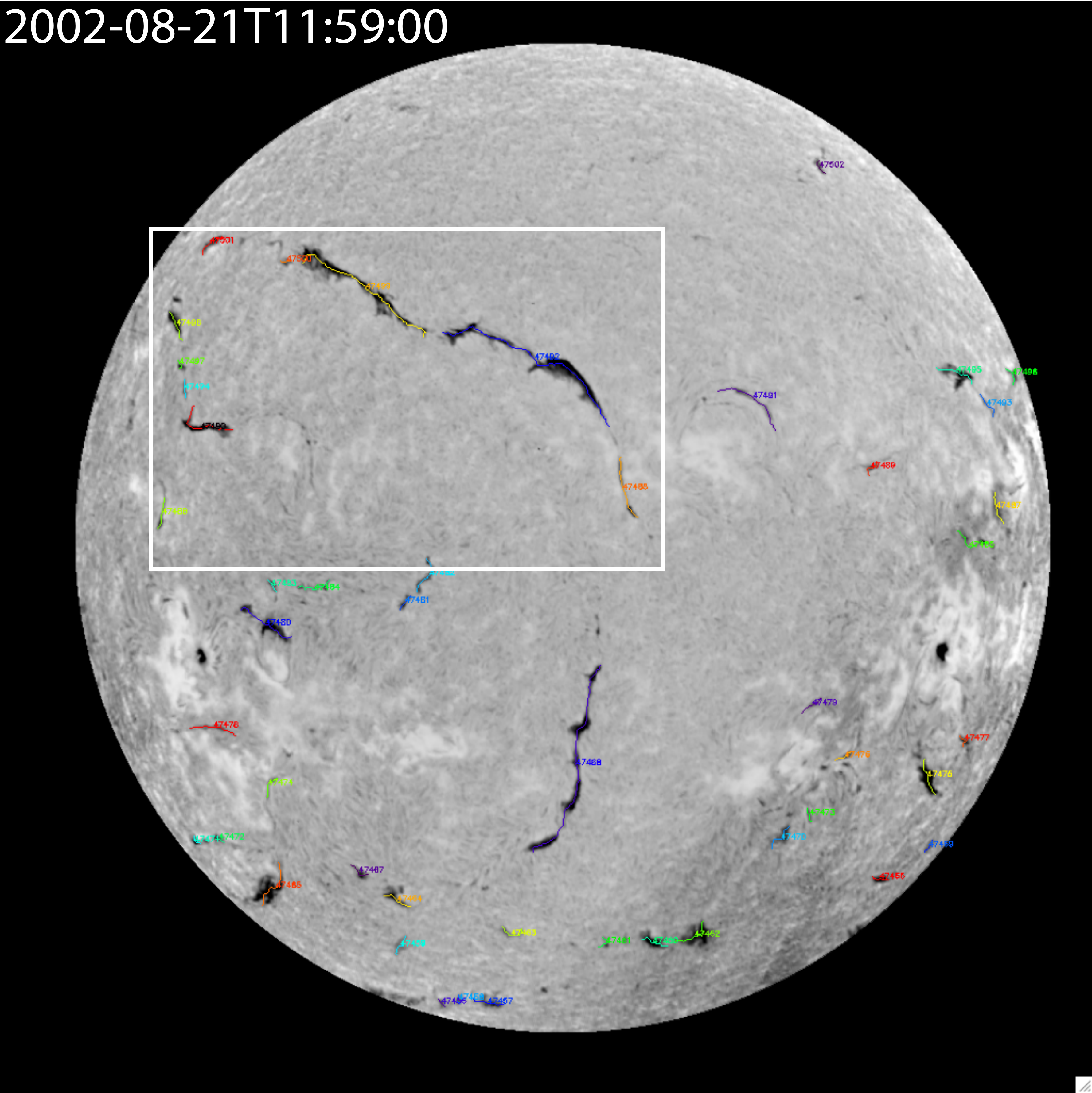} d.\includegraphics[width=0.50\textwidth]{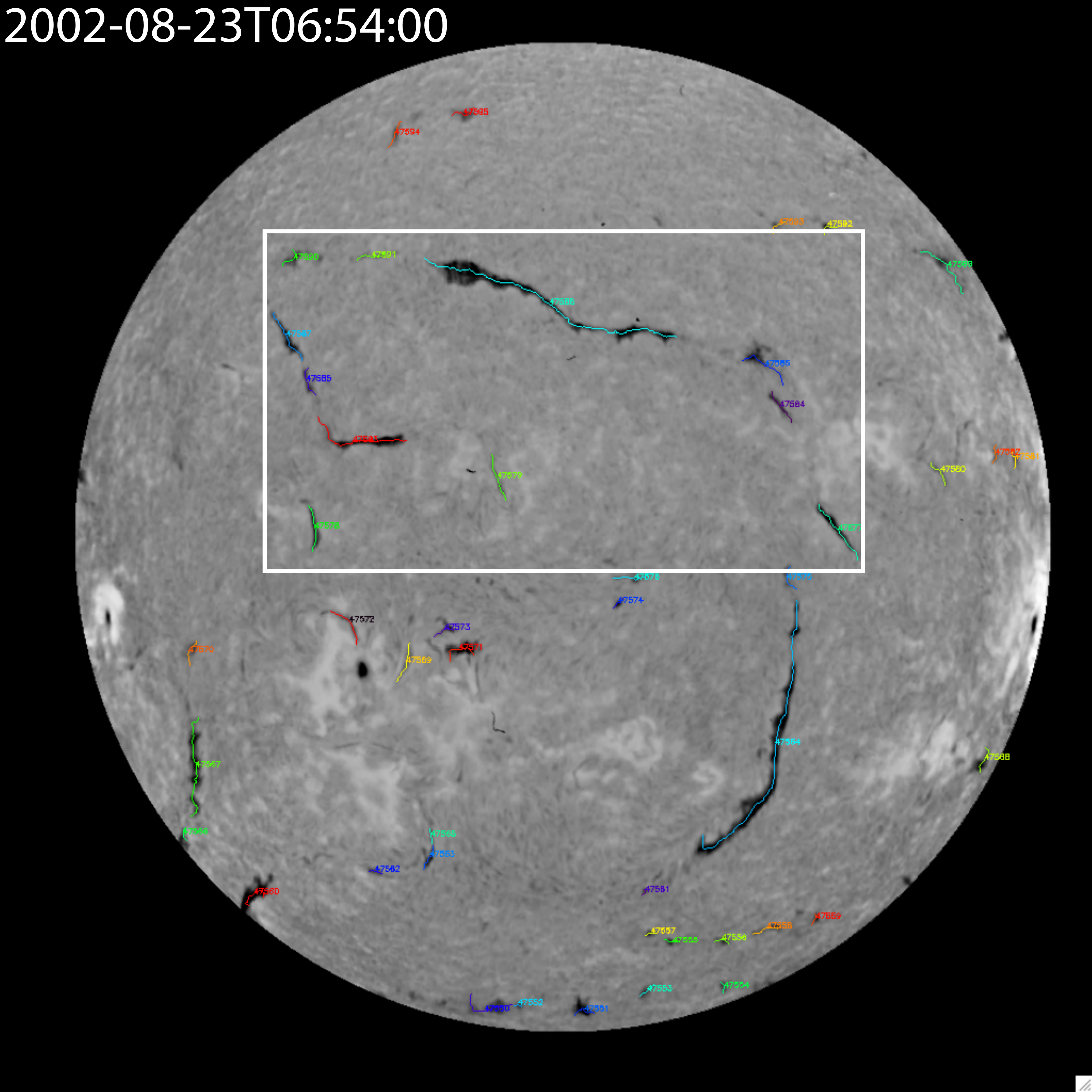}}
  \centerline {e.\includegraphics[width=0.50\textwidth]{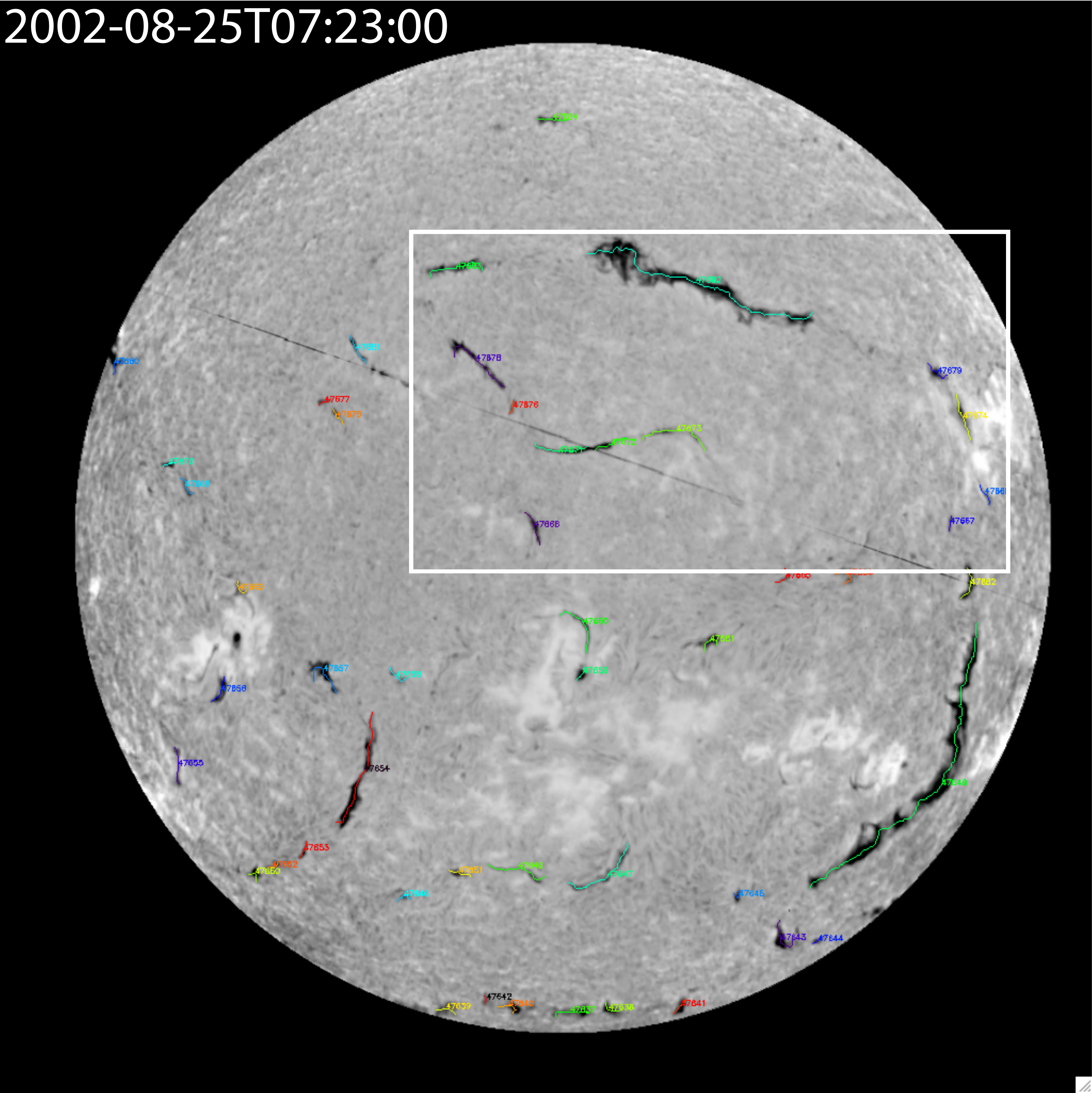} f.\includegraphics[width=0.50\textwidth,height=6.0cm]{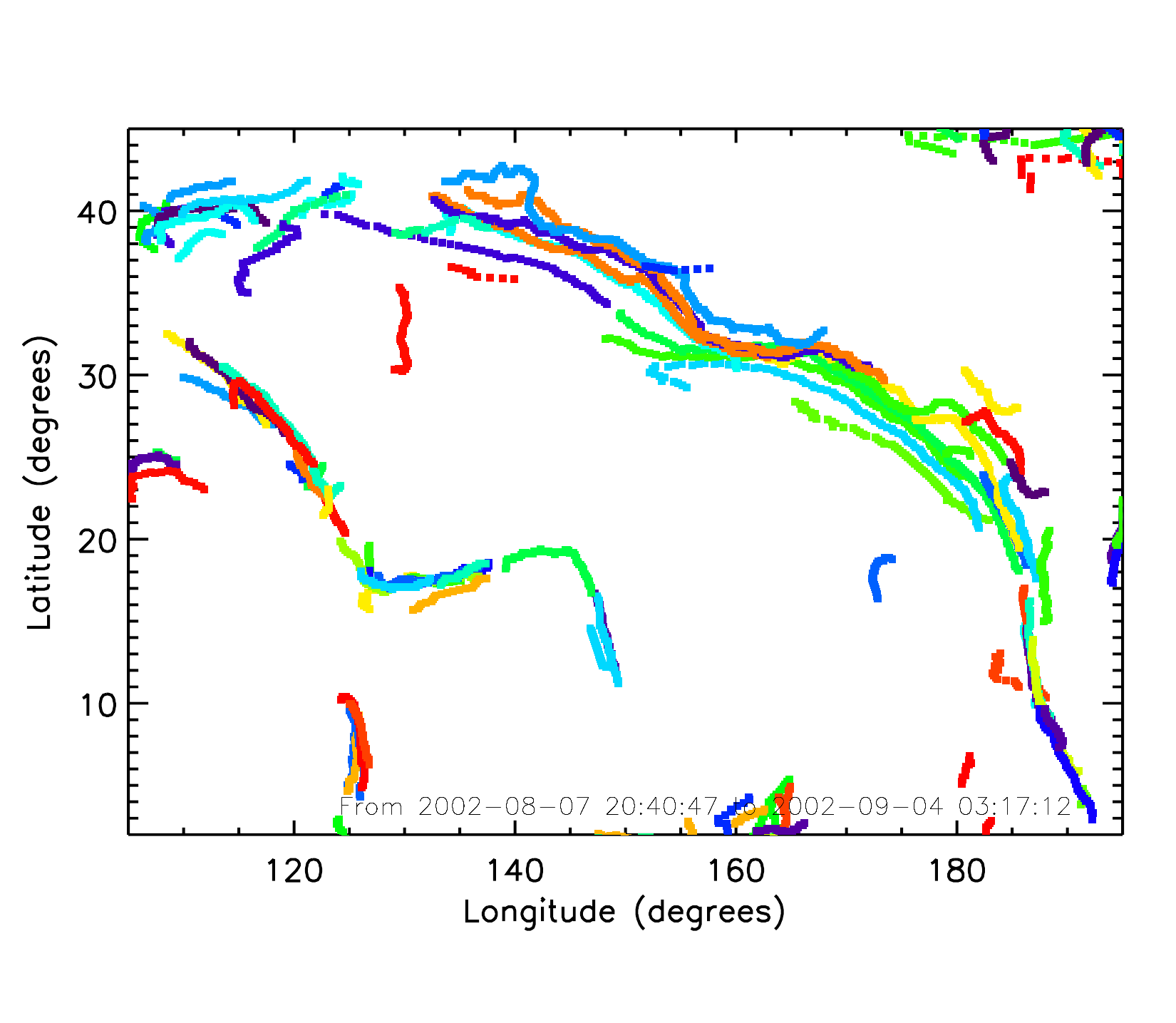}}
  \caption{Illustration of the calculation of the skeleton pixel's coordinates from the heliocentric coordinates system to the co-rotating one.  Panels a, b, c, d, and e show a sample of five images of the Sun produced between 7 August 2002 at 20:40:47 UT and 4 September 2002 at 03:17:12 UT (on each image the date of observation is written in the upper-left corner). Example of skeleton pixel's coordinates $(\varphi,\lambda)$ calculated in the co-rotating frame $\Re$ are plotted on Panel f; the region displayed corresponds to the white bounding rectangles on each image.}\label{fig:mh2re}
\end{figure}

The steps of calculation of the skeleton coordinates in the reference frame $\Re$ are illustrated on Figure \ref{fig:mh2re}. In this example, we want to track filaments detected on a Meudon spectroheliogram on 21 August 2002 at 11:59:00 UT; date that defines here the time of observation $[t_{\iota}]$. Thus, according to the discussion in Section \ref{S:Algorithm}, the tracking process concerns all of the filaments extracted between 7 August 2002 at 20:40:47 UT and 4 September 2002 at 03:17:12 UT, which correspond to the start time $[t_0]$ and the end time $[t_1]$ of the tracking period respectively. Panels a, b, c, d, and e, show a sample of five images of the Sun obtained during this time range; the date of observation is indicated in the upper-left corner of each image (for more clarity only a few images produced during this period are displayed). The heliocentric coordinates of skeleton pixels as well as the identification numbers of each detected filament are also plotted using a coloured scale. Panel f presents an example of corresponding coordinates $(\varphi,\lambda)$, calculated for the co-rotating filament visible inside the white bounding rectangles on the previous panels. In $\Re$, all of the skeletons belonging to this filament form a structured cluster around the same location, giving the full shape of the filament over the disk crossing.

\begin{figure} 
  \centerline{\includegraphics[width=1.00\textwidth]{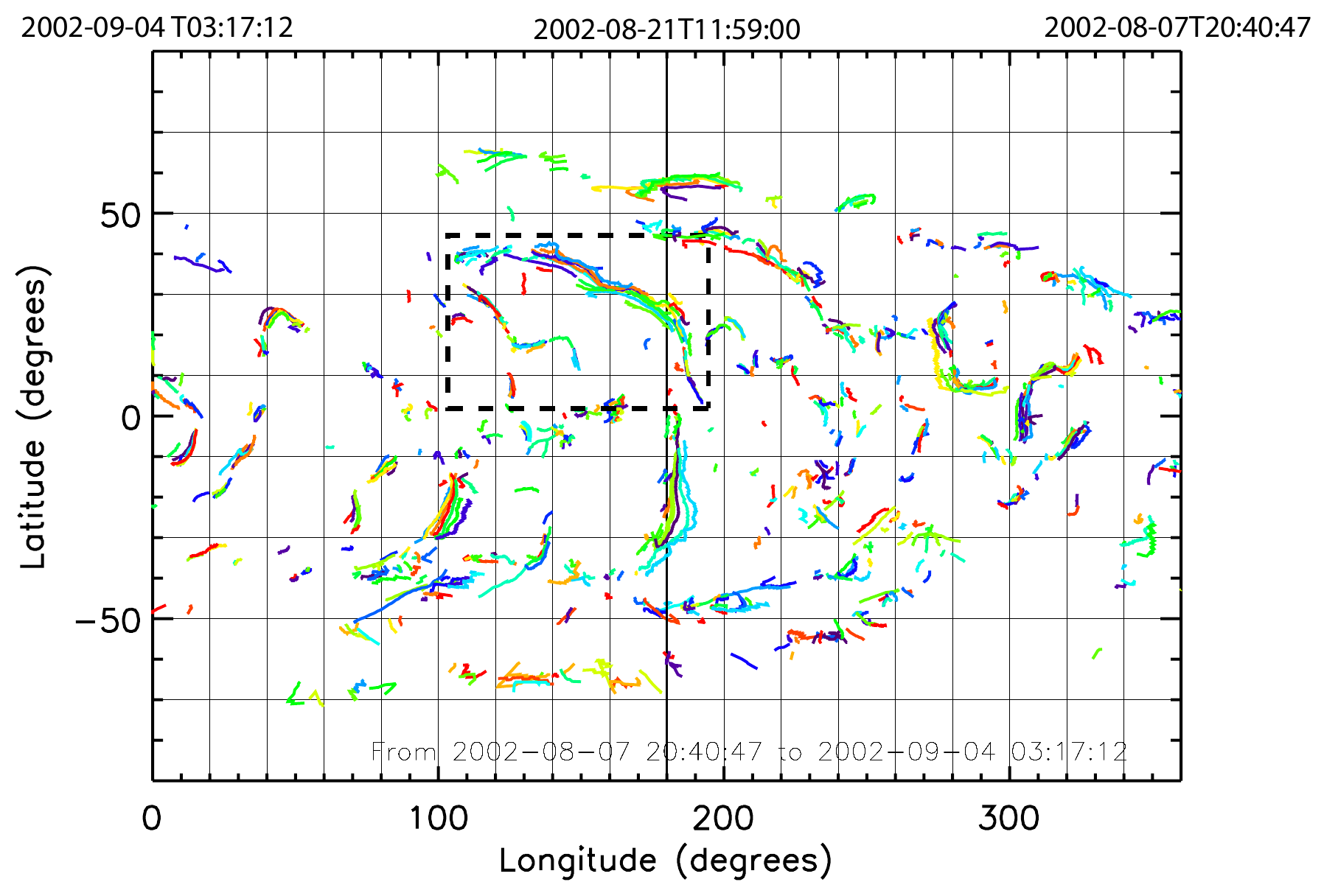}}
  \caption{Synoptic map of the skeleton pixel's coordinates in $\Re$ between 07 August 2002 at 20:40:47 UT (\textit{i.e.}, $t_0$) and 04 September 2002 at 03:17:12 UT (\textit{i.e.}, $t_1$). Carrington convention is used to define the longitude axis. The vertical darker line at the centre indicates the longitude $\varphi_{t_{\iota}} = 180^\circ$, which corresponds the longitude of the central meridian on 21 August 2002 at 11:59:00 UT. The bounding rectangle delimited by black dashed line shows the area displayed on the panel f in Figure \ref{fig:mh2re}.}\label{fig:SynopticMap}
\end{figure}

Figure \ref{fig:SynopticMap} presents the resulting synoptic map of skeleton pixel's coordinates converted in $\Re$ for the previous example over the full period $[t_0,t_1]$. The longitude axis uses the Carrington convention starting at $\varphi_{t_{0}} = 360^\circ$, and ending at $\varphi_{t_{1}} = 0^\circ$, with $\varphi_{t_{\iota}} = 180^\circ$, corresponding to the longitude of the central meridian on 21 August 2002 at 11:59:00 UT. At this stage of the procedure, the heliocentric and co-rotating pixel coordinates of features, as well as their corresponding identification numbers $[\nu$ and $\tau]$, are stored by the tracking code.

\subsubsection{Curvilinear Interpolation} 
  \label{S:Interpolation}
  
To be consistent, the curve-matching algorithm must be applied on two curve segments having the same number of points and the same length. However, due at least to the coordinates' projections, the distribution of points along the skeleton curves is not necessarily uniform, hence, we need to interpolate the curves in such a way that the distance $[ds]$ between two consecutive points is constant. At the same time, such a procedure can be seen as a smoothing operation that will highlight the main shape of the skeletons. It results that the value in degrees of $[ds]$ to use in the co-rotating frame has to be a compromise between the spatial resolution (\textit{i.e.}, minimum distance between two points) and the length of the skeletons: too small a value can significantly increase the computation time, but also the possible spatial fluctuations along the curve at small scales, whereas too large a value decreases the number of points along the curve, and so the spatial resolution used to define its shape.\\
Once an acceptable value of $[ds]$ is found, starting from the location of the first point $(\varphi_{1},\lambda_{1})$ at one of the ends of the curve (the order of the points along the curve goes from one end to the other), the interpolation method consists in the determination of the intersection point $(\varphi_{k_{int}},\lambda_{k_{int}})$ between the circle of radius $[ds]$ centred on $(\varphi_{1},\lambda_{1})$, and the line passing through the two points $(\varphi_{k},\lambda_{k})$ and $(\varphi_{k+1},\lambda_{k+1})$ of the curve, which satisfy the condition:
\begin{equation}\label{eq:ds}
\sum_{l=2}^{k}\sqrt{(\varphi_{l} - \varphi_{l-1})^2 + (\lambda_{l} - \lambda_{l-1})^2}\leqslant ds \leqslant \sum_{l=2}^{k+1}\sqrt{(\varphi_{l} - \varphi_{l-1})^2 + (\lambda_{l} - \lambda_{l-1})^2}.
\end{equation}  
When the intersection point $(\varphi_{k_\mathrm{int}},\lambda_{k_\mathrm{int}})$ is known, the process is then repeated replacing $(\varphi_{1},\lambda_{1})$ by $(\varphi_{k_\mathrm{int}},\lambda_{k_\mathrm{int}})$ as the new circle centre, and calculating the next intersection point $(\varphi_{k_\mathrm{int+1}},\lambda_{k_\mathrm{int+1}})$ with the curve. The interpolation finally stops when the other end of the curve is reached.\\
At the same time, the lengths of the skeletons are estimated by simply summing the distances $[ds]$ found between each interpolated points (\textit{i.e.}, $(\varphi_{k_\mathrm{int}},\lambda_{k_\mathrm{int}})$ and $(\varphi_{k_\mathrm{int+1}},\lambda_{k_\mathrm{int+1}})$). This technique will slightly under-estimated the actual length, which appears to be not really significant on the tracking results.

\subsubsection{Curve Matching Algorithm} 
  \label{S:CMA}
 
In order to significantly reduce the computation time, the matching algorithm is only applied on skeletons that are close enough on the reference frame. To achieve this goal, for each coordinates $(\varphi_{k_i},\lambda_{k_i})$ of a first skeleton $i$, we search the coordinates $(\varphi_{k_j},\lambda_{k_j})$ of others skeletons $j$ ($j \neq i$) for which the distance $r_{k_ik_j}=\sqrt{(\varphi_{k_i} - \varphi_{k_j})^2 + (\lambda_{k_i} - \lambda_{k_j})^2}$ is less than a maximum value $r_\mathrm{max}$ (defined in degrees in $\Re$). If the condition $r_{k_ik_j}<r_\mathrm{max}$ is fulfilled, then the curve matching algorithm is applied. In practice the value of the input parameter $r_\mathrm{max}$ does not affect significantly the efficiency of the algorithm, because it will automatically dissociate features that are too much distant.\\
 
Given two curves $C_1$ and $C_2$ of close skeletons $1$ and $2$ from the same filament observed on two different images at two different times, there should be a Euclidean transformation $E$, such that $EC_1$ matches $C_2$. However, since during its lifetime a filament can be subject to shape deformation, segmentation, but also translation and/or rotation, no exact match is likely to occur, and so we look for a transformation $E$ giving the best match in the least-squares sense. Let $C_1$ and $C_2$ be represented by the vectorial sequences $(\mathbf{u_k})^{n}_{k=1}$ and $(\mathbf{v_k})^{n}_{k=1}$  respectively, where $n$ is the number of points of the curves (we assume here that the two curves have the same length). Matching consists of finding a Euclidean transformation $E$ of the plane that will minimize the distance $[l^2]$ between the sequences $(E\mathbf{u_k})^{n}_{k=1}$ and $(\mathbf{v_k})^{n}_{k=1}$, such that:

\begin{equation}\label{eq:delta}
\Delta = \mathrm{min}_E l^2= \mathrm{min}_E \sum_{k=1}^{n}\vert E\mathbf{u_k} - \mathbf{v_k} \vert ^2.
\end{equation}
To simplify the calculation, the curve $C_1$ is first translated such as $\sum_{k=1}^{n} \mathbf{u_k} =0$. Then, if we write $E$ as $E\mathbf{u_k} = R_{\mathbf{\theta}}\mathbf{u_k} + \mathbf{a}$, where $R_{\theta}$ denotes a counter-clockwise rotation by an angle $\mathbf{\theta}$, and $\mathbf{a}$ a translation; the best match is obtained when \cite{AyacheIEEE1986,WolfsonIEEE1990}
\begin{equation} 
\mathbf{a} = \frac{1}{n} \sum_{k=1}^{n} \mathbf{v_k}\mbox{, and }\mathbf{\theta} = -\sum_{k=1}^{n}u_k\overline{v_k},
\end{equation}
where $u_k$ and $\overline{v_k}$ are complex representations of the vectors $\mathbf{u_k}$ and $\mathbf{v_k}$ respectively.
\linebreak 

At the end, for each pair $(C_i,C_j)$ of skeleton curves analysed, the matching process returns three best-fitted parameters 
\begin{equation}
\left( a = \vert \mathbf{a} \vert + \delta a, \theta = \vert \mathbf{\theta} \vert,l=\sqrt{\Delta}\right)_{ij},
\end{equation}
where $\delta a$ is a term that corrects the $\sum_{k=1}^{n} \mathbf{u_k} =0$ translation introduced previously. These parameters, which represent respectively the degrees of translation, rotation, and deformation between the two curves, will be used to compute in a second step the probability $[P_{ij}]$ that the skeletons $[i$ and $j]$ belong to the same filament. \\
If two skeletons have different lengths, $l_i \neq l_j$, the matching is also performed but between the smallest of the two, say $i$, and all of the possible sub-segments $k$ of the other one $j$ that have the same length $l_k = l_i$. In this case, only the probability $[P_{ik}]$ for the sub-segment returning the best match is finally retained. This technique becomes actually consistent only if the lengths of the two curves do not differ significantly, we therefore introduce an additional condition to the minimum on maximum lengths ratio: if this ratio is too small, then the matching algorithm is not applied.

\subsubsection{Confidence of Tracking} 
  \label{S:Probability}

\begin{figure} 
  \centerline{\includegraphics[width=0.5\textwidth]{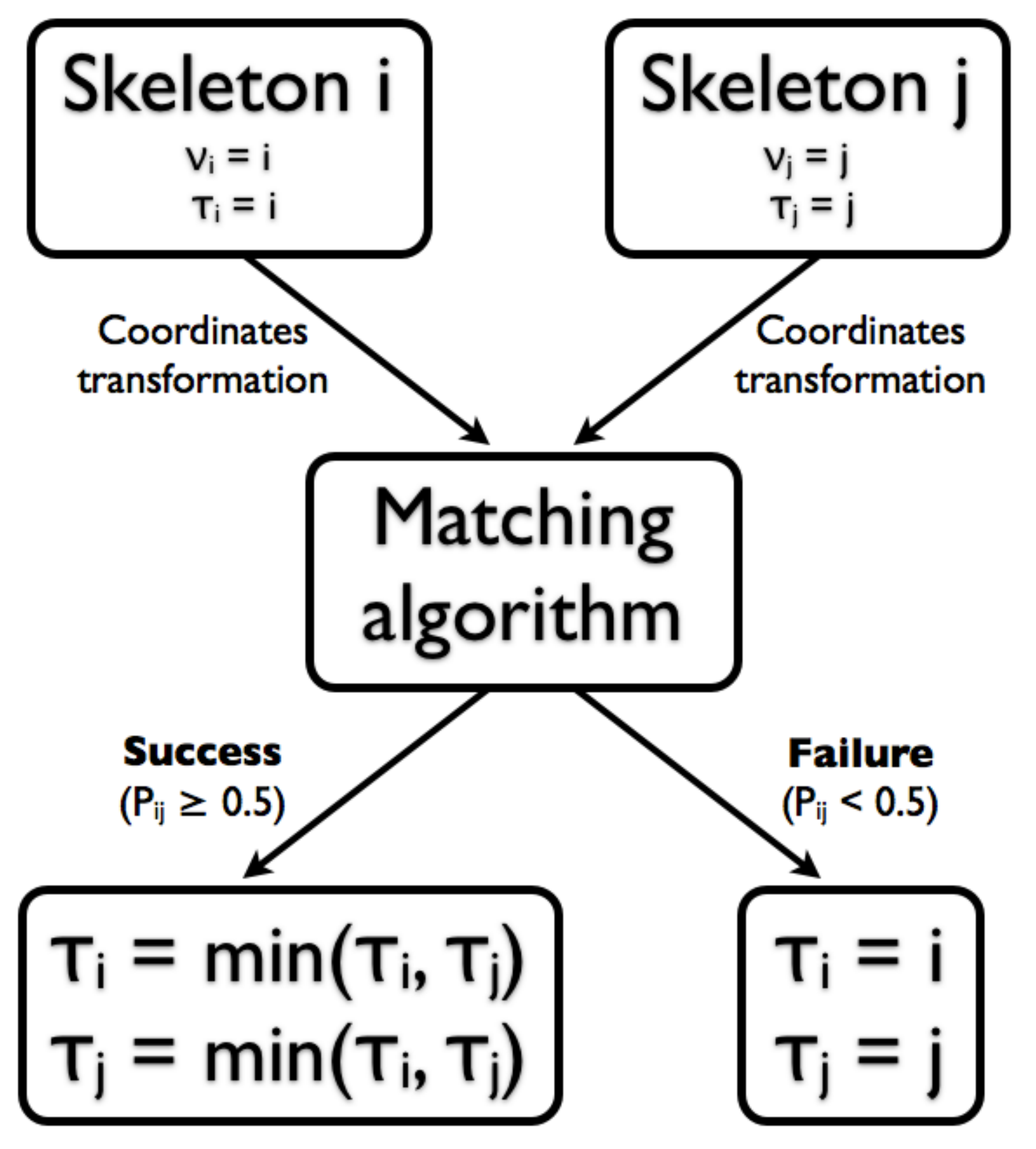}}
  \caption{Given two skeletons $[i$ and $j]$ of feature and tracking identification numbers initially equal to $(\nu_i = i, \tau_i = i)$ and $(\nu_j = j, \tau_j = j)$ respectively, where $i \neq j$, the association consists of assigning the same tracking identification number to both skeletons when the matching succeeds. In practice the minimum value between $\tau_i$ and $\tau_j$ is conserved. In the case where at least one of the skeletons compared has already received a different tracking number, from a previous matching with another skeleton $k$ for instance, $\tau_i = \tau_k \neq i$, then we assign to all the features $[i$, $j$, and $k]$ the same number $\tau_i = \tau_j = \tau_k = \mathrm{min}(\tau_i,\tau_j,\tau_k)$.}\label{fig:Diagram}
\end{figure}

For each set of best-fitted parameters $(a,\theta,l)_{ij}$, we calculate a normalized probability $[P_{ij}]$ defined by: 
\begin{equation}\label{eq:P}
P_{ij} = W_a P_a + W_{\theta} P_{\theta} + W_l P_l\quad (0 \leq P_{ij} \leq 1),
\end{equation}
where $P_a$, $P_{\theta}$, and $P_l$ are the probability terms respectively related to the translation $[a]$, rotation $[\theta]$, and deformation $[l]$ parameters, and $W_a$, $W_{\theta}$, and $W_l$ are the corresponding probability weights, which satisfy $W_a + W_{\theta} + W_l = 1$. $P_{ij}$ can be seen as the level of trust of the tracking; the larger it is, the more the automated code is confident about the fact the two skeletons $[i$ and $j]$ are actually two segments of the same filament.\\  
Since no formal relation exists, the $P_a$, $P_{\theta}$, and $P_l$ terms are simply described here using linear expressions, nevertheless, under the condition that the probability $[P_{ij}]$ decreases when the three parameters $a$, $\theta$, and $l$ increase:
\begin{eqnarray}\label{eq:Plin}
P_a = 1 - \frac{a}{a_0}(1-P_{\Theta})\mbox{ if }a \leq \frac{a_0}{1-P_{\Theta}}\mbox{, and }P_a = 0\mbox{ otherwise,}\\
P_{\theta} = 1 - \frac{\theta}{\theta_{0}}(1-P_{\Theta})\mbox{ if }\theta \leq \frac{\theta_{0}}{1-P_{\Theta}}\mbox{, and }P_{\theta} = 0\mbox{ otherwise,}\\
P_l = 1 - \frac{l}{l_0}(1-P_{\Theta})\mbox{ if }l \leq \frac{l_0}{1-P_{\Theta}}\mbox{, and }P_l = 0\mbox{ otherwise,}
\end{eqnarray}
where $a_0$, $\theta_{0}$, and $l_0$ are input parameters that satisfy $P_{a=a_0} = P_{\Theta}$, $P_{\theta=\theta_0} = P_{\Theta}$, and $P_{l=l_0} = P_{\Theta}$) respectively. At the same time, the three conditions $a \leq \frac{a_0}{1-P_{\Theta}}$, $\theta \leq \frac{\theta_{0}}{1-P_{\Theta}}$, and $l \leq \frac{l_0}{1-P_{\Theta}}$ will ensure that $P_a$, $P_{\theta}$, and $P_l$ have always positive values. The determination of these three parameters' values will be explained later in Section \ref{S:Results}. Concerning $P_{\Theta}$, we take $P_{\Theta} = 0.5$ which is actually also the value of the respective probability $[P_{ij}]$ above which two skeletons are assumed to match, and are thus associated. \\
The association step is very important since it will permit us to follow a filament on successive images, assigning the same tracking identification number to its skeletons. An example of such a process is described on Figure \ref{fig:Diagram}. Where two skeletons $[i$ and $j]$ are not associated together (\textit{i.e.}, $\tau_i \neq \tau_j$), but both are associated with a third one $[k]$ (\textit{i.e.}, $\tau_i = \tau_k$ and $\tau_j = \tau_k$), then all three will be related, and will obtain the same tracking identification number (i.e., $\tau_i = \tau_j = \tau_k$). This condition appears to be very efficient for merging several segments of a filament that suffers significant shape modifications.

Once all of the associations between the pairs of curves are done, for each skeleton $[i]$ we finally calculate the average value
\begin{equation}\label{eq:Pij}
<P_{i}> = \frac{1}{N_j}\sum_{j=1}^{N_j} P_{ij}, 
\end{equation}
where $N_j$ is the number of matchings realized between the skeleton $[i]$ and the skeletons $[j]$ ($j \neq i$) for which $\tau_j = \tau_i$. This value will be used as an average tracking confidence level for the corresponding filament $[i]$.   

\subsection{Tracking Filaments over Two Successive Solar Rotations} 
  \label{S:SeveralRotations}
  
\begin{figure} 
  \centerline{\includegraphics[width=1.00\textwidth]{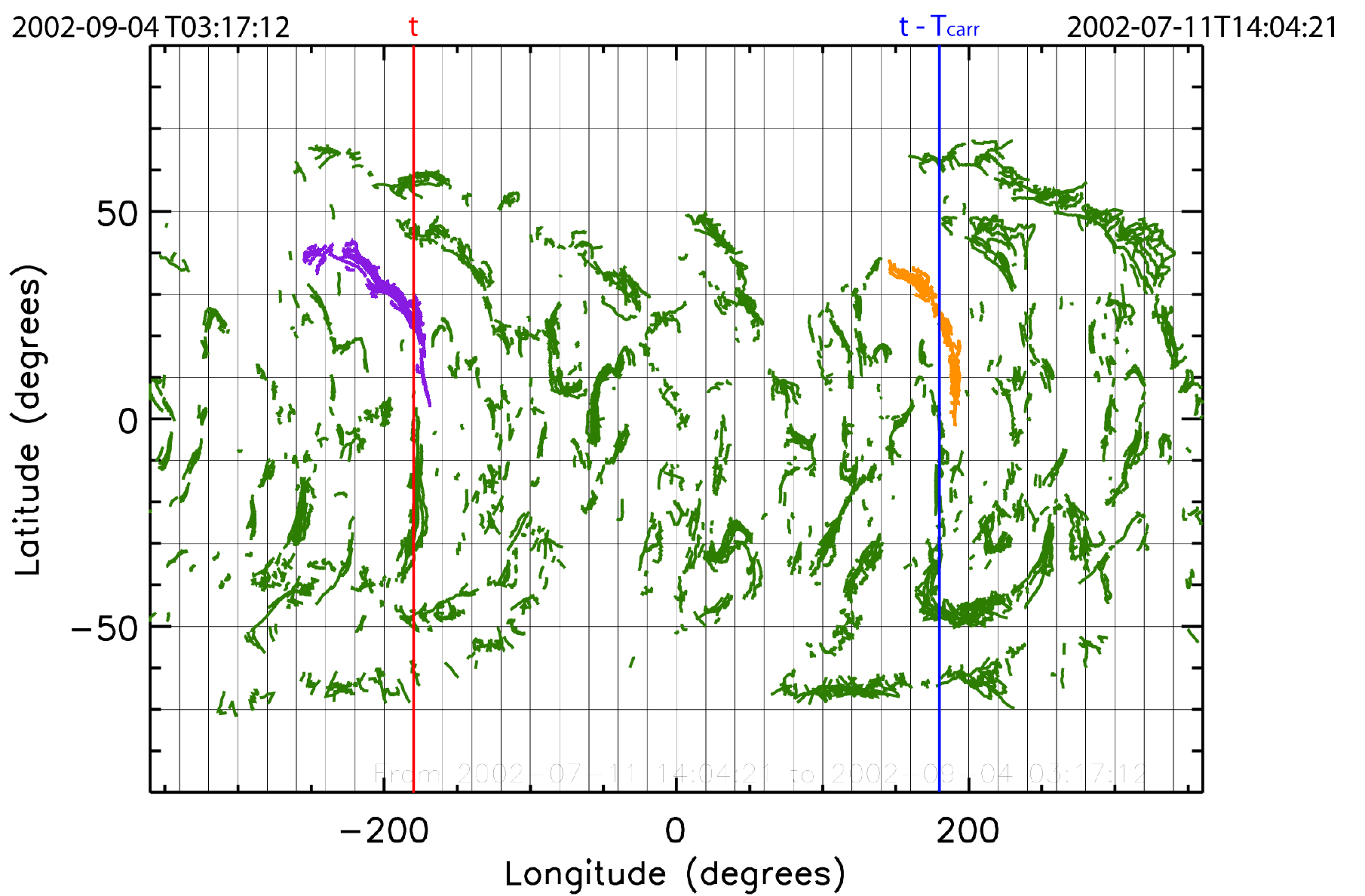}}
  \caption{Synoptic map of the new skeleton pixel's coordinates between $t_0$ at 14:04:21 UT on 11 July 2002, and $t_1$ at 03:17:12 UT on 4 September 2002. Carrington convention is still used here to define the longitude axis. The vertical red line on the left indicates the longitude $\varphi_{t} = -180^{\circ}$, which corresponds the longitude of the central meridian at $t$, on 21 August 2002 at 11:59:00 UT. The vertical blue line on the right indicates the longitude $\varphi_{t-T_\mathrm{carr}} = +180^{\circ}$, which coincides with the longitude of the central meridian at $t-T_\mathrm{carr}$. The two groups of skeletons coloured respectively in purple and orange belong to the same filament observed over two solar rotations (all the others skeletons are plotted here in green for more clarity). In this example, the matching process is first used to group together the purple skeletons on one side, and the orange skeletons on the other. Then, when each group has its own tracking identification number, the matching algorithm is run to compare the curves of the two clusters using Equation (\ref{eq:Pp}). In this case the matching process can be done because both group gravity centres are separated by $360^{\circ}$.}\label{fig:SynopticMap2}
\end{figure}

The algorithm can also be adapted to track filaments that perform two consecutive crossings on the solar disk. If we assume that a given filament is observed around a time $t_{\iota}$, but also around $t_{\iota} - T_{\mathrm{carr}}$, then we can extend the previous time range $[t_0,t_1]$ such that $t_0 \approx t_{\iota} - 1.5T_{\mathrm{carr}}$ and $t_1 \approx t_{\iota} + 0.5T_{\mathrm{carr}}$, where the average coordinates of the filament skeletons in the current and in the previous rotations will be respectively $(<\!\varphi\!>,<\!\lambda\!>)$ and $(<\!\varphi\!> + 360^{\circ},<\!\lambda\!>)$ (using the reference convention defined by Equation (\ref{eq:coord1}) and Equation (\ref{eq:coord2})). Figure \ref{fig:SynopticMap2} shows the skeleton pixel's coordinates in $\Re$ between $t_0$ and $t_1$, taken $t_{\iota}$ at 11:59:00 UT on 21 August 2002.\\
At this stage, we first apply the matching algorithm in order to set the tracking identification numbers (as described on previous sections), then from a first group $[g_i]$ of skeletons $[i]$ having the same tracking identification number $\tau_i = \tau_{g_i}$ and located around $(<\!\varphi_i\!>,<\!\lambda_i\!>)$, we look for a possible group $[g_j]$ of skeletons $[j]$, satisfying $\tau_j = \tau_{g_j}$ and situated around the coordinates $(<\!\varphi_j\!> \approx <\!\varphi_i\!> + 360^{\circ},<\!\lambda_j\!> \approx <\!\lambda_i\!>)$. If such a group $[g_j]$ is found, then the matching algorithm is run between the skeletons of both groups only. In practice, to start the matching process, the average gravity center $(<\!\varphi_j\!>,<\!\lambda_j\!>)$ of the group $[g_j]$ must be inside the circle of radius $r_\mathrm{max}$ and centred on $(<\!\varphi_i\!> + 360^{\circ},<\!\lambda_i\!>)$.

However, since the chance to observe the same filament over several successive rotation periods decreases with time, we introduce here a new probability term $[P_{\Delta t}]$ that is proportional to the difference $\Delta t$ of the observation times between the two skeletons to match. (Since one solar rotation period separates the two features, $\Delta t$ should be normally around $T_\mathrm{carr}$.) Hence, the resulting probability $[P'_{ij}]$ that we calculate for each pair $(i,j)$ of skeletons of the two groups will be slightly different:

\begin{equation}\label{eq:Pp}
P'_{ij} = P_{\Delta t} P_{ij},
\end{equation}
 
\noindent where $P_{ij}$ is the probability defined in Equation (\ref{eq:P}).\\  
In order to reproduce the filament's average lifetime, the term $P_{\Delta t}$ must be maximal (\textit{i.e.}, $P_{\Delta t} = 1$) when $\Delta t = 0$, and must tend towards zero when $\Delta t$ increases. As for $P_a$, $P_{\theta}$, and $P_l$ (see Equation (\ref{eq:Plin})), we use here a linear expression defined by:

\begin{equation}\label{eq:Pt}
P_{\Delta t} = 1 - \frac{\Delta t}{\Delta t_0}(1-P_{\Theta})\mbox{ if }\Delta t \leq \frac{\Delta t_0}{1-P_{\Theta}}\mbox{, and }P_{\Delta t} = 0\mbox{ otherwise,}
\end{equation}

\noindent where $\Delta t_0$ is an input parameter above which the probability $[P_{\Delta t}]$ falls below a given value $P_{\Theta} = 0.5$. The value of $\Delta t_0$ is also estimated empirically as explained in the next section.\\
Finally, the code will associate the two groups $[g_i$ and $g_j]$ of skeletons only if the average value 
\begin{equation}\label{eq:Pp}
<\!P'\!> = \frac{1}{2}\frac{1}{N_i}\frac{1}{N_j}\sum_{i=1}^{N_i}\sum_{j=1}^{N_j} P'_{ij}, 
\end{equation}
where $N_i$ and $N_j$ are respectively the number of skeletons in the first and second group, is greater then $0.5$. In this case, all of the skeletons $[i]$ of the first group $[g_i]$ will receive a third identification number $[\omega_i]$, which will be equal to the tracking identification number of the group $[g_j]$ such as $\omega_i =\tau_j = \tau_{g_j}$. \\
At the end of the tracking computation, a filament will be specified by three identification numbers $[\nu_i$, $\tau_i$, and $\omega_i]$ to be identified on a image, to be tracked over the solar disk, and to be possibly linked to another feature on the previous rotation respectively. This information will be written in a dedicated tracking table of the HFC.

\subsection{Filaments Behaviour} 
  \label{S:FilamentBehaviour}  
 
In addition to the tracking algorithm, a module has been implemented in the code to characterize the behaviour of the filaments during their crossing the solar disk. To achieve this goal, the procedure checks, using the tracking identification numbers, that each co-rotating filament is well detected on every image between its first and last times of observation. If during this period the filament disappears (\textit{i.e.}, is not seen on one or more successive images), then the times corresponding to the last observation before its disappearance and the first observation after its re-appearance are saved. Moreover, if the filament appears once it has crossed the east limb, and/or disappears before reaching the west limb, the information is also returned by the code. To proceed in this case, average heliographic coordinates $(<\!\varphi_{h}\!>,<\!\lambda_{h}\!>$ of the filament skeleton centres on the first/last image are used to calculate the predicted longitudes on the observation just before/after; if these longitudes are less than $70^{\circ}$ (in absolute value), then we assume that there is an appearance/disappearance after/before the limb. As for the tracking data, the results of the analysis are stored in the tracking table of the HFC.
  
\section{Performances of the Algorithm} 
  \label{S:Performances}
 
\subsection{Assessments}
	\label{S:Assessments}

To optimize the tracking code, we need to define the combination of input parameters $[a_0,\theta_{0},l_{0},\Delta t_{0}]$ that gives the best matching rate. This set will serve as a reference for tracking the filaments on the Meudon H$\alpha$ spectroheliograms. At the same time, the assessment of these parameters will also give an indication of the performance of the algorithm, providing a \textit{positive} tracking rate $[R_{p}]$ after each run (as explained below). \\
To find the best set of input parameters, we compared the code results with a representative sample of filaments manually tracked. This sample contains approximately 2000 filaments, detected by the recognition code between 2000 and 2009 on Meudon spectroheliograms. The tracking was then performed identifying "by-hand" the co-rotating filaments on synthetic synoptic maps, such as the one displayed on Figure \ref{fig:SynopticMap}. Direct checking on corresponding H$\alpha$ images was also done to confirm the first choice. In order to keep a track of this selection, the results are written in ASCII format files using identification numbers $(\nu_{i}^\mathrm{man},\tau_{i}^\mathrm{man},\omega_{i}^\mathrm{man})$ as in the automatic code. In practice, since all filaments were loaded from the HFC, the identification numbers $[\nu_{i}^\mathrm{man}$ and $\nu_{i}]$ are actually identical; this criterion will permit us to identify individual features between both sets of results.\\
Once the sample was generated, we then proceeded as follows:

\begin{enumerate}
\item The values of the input parameters $ds$, $r_\mathrm{max}$, $W_a$, $W_{\theta}$, $W_l$, $P_{\Theta}$, and $\Theta$ were first fixed to a given value. (The effects of these parameters on the code performance were also tested, but the results of this evaluation are not presented here to not overload the discussion.). We took respectively, $ds = 1^{\circ}$, $r_\mathrm{max}=5^{\circ}$, $W_a = W_{\theta} = W_l = 1/3$, $P_{\Theta} = \Theta = 0.5$, and the ratio of minimum to maximum lengths of the compared filaments had to be greater than $1/3$. Since the application of the tracking algorithm to too small filaments is neither relevant nor efficient, we thus decided to process only the filaments for which the length is equal or greater than $5^{\circ}$.
\item We carried out a series of executions of the code, using different combinations of values for the input parameters $a_0$, $\theta_{0}$, $l_{0}$, and $\Delta t_0$ (reasonable ranges of values were previously defined for each parameter). 
After each computation, a \textit{positive} tracking rate $[R_{p}]$ was systematically calculated to evaluate the performance of the automated process compared to the manual one. In order to estimate $R_{p}$, we compare two by two the groups of filaments from both manual and automatic sets, which can be identified by their own tracking identification numbers $\tau_{i}^\mathrm{man}$ and $\tau_{i}$ respectively. For two filaments $[i$ and $j]$ belonging to each type of groups, if the condition $\nu_{i}^\mathrm{man}$=$\nu_{j}$ is fulfilled, then $R_{p}$ was incremented by $1/n$ where $n$ is the total number of comparisons. 
\item Finally, we keep the combination of values of $a_0$, $\theta_{0}$, $l_{0}$, and $\Delta t_{0}$ that gives the highest \textit{positive} rate. 
\end{enumerate}

In the first computation series, we did not take account of the results of the tracking over two rotations including $\omega_{i}$, then in a second series, we performed the same operation including both $\tau_{i}^\mathrm{man}$ and $\omega_{i}^\mathrm{man}$ to compute $R_{p}$.
Results of the computed series are presented in the following section.
   
\subsection{Results}   
		\label{S:Results}
   
Table \ref{tbl:results1} shows the results of a first series of comparisons between the automated and the manual processes. As explained in the previous section, this first series does not take account of the tracking from one rotation to another using $\omega_{i}^\mathrm{man}$. The values of the input parameters given the highest rate $R_{p}$ are in this case: $a_0 = 5^{\circ}$, $\theta_0 = 30^{\circ}$, and $l_0 = 4^{\circ}$. $R_{p}$ and the corresponding numbers of tracked filaments, which have a length of skeleton $[l_\mathrm{ske}]$ greater than $5^{\circ}$, $6^{\circ}$, $8^{\circ}$, $10^{\circ}$, and $15^{\circ}$ respectively, are indicated. ($R_p$ is provided in percent.)\\ 
According to Table \ref{tbl:results1}, the performance of the code increases when the length $l_\mathrm{ske}$ increases: passing from $86\%$ of successful tracking for the $N_\mathrm{fil}=1991$ filaments that have a length greater than $5^{\circ}$, to $95\%$ for the $N_\mathrm{fil}=479$ filaments for which $l_\mathrm{ske} \geq 15^{\circ}$. This is the consequence of the algorithm sensitivity to the length of the skeletons, since it is more difficult for the curve-matching to compare filaments that have too small lengths.
 
\begin{table}
\caption{Results of the assessments for the tracking over one crossing on the solar disk: the \textit{positive} rate $R_{p}$ in percent (second line) and the corresponding number of filaments $N_\mathrm{fil}$ (third line), using the best input parameters, are indicated for different lengths of filament $l_\mathrm{ske}$ (first line).}\label{tbl:results1}
\begin{tabular}{cccccc}     
\hline
\hline
$l_\mathrm{ske}$ & $\geq 5^{\circ}$ & $\geq 6^{\circ}$ & $\geq 8^{\circ}$ & $\geq 10^{\circ}$ & $\geq 15^{\circ}$ \\
\hline
$R_{p}$ & $86\%$ & $89\%$ & $92\%$ & $94\%$ & $95\%$ \\
$N_\mathrm{fil}$ & $1991$ & $1698$ & $1263$ & $924$ & $479$ \\
\hline
\end{tabular}
\end{table}  

Table \ref{tbl:results2} shows the results of the second series of assessment, including here the tracking results from a rotation to the following.  In this case, the input parameters that give the best \textit{positive} rate $R_{p}$ are : $a_0 = 5^{\circ}$, $\theta_0 = 20^{\circ}$, $l_0 = 2^{\circ}$, and $\Delta t_0 = 112$ decimal days. The total \textit{positive} rate $[R_{p}]$ (given in percent) and the corresponding numbers of tracked filaments, which have a length of skeleton $[l_\mathrm{ske}]$ greater than $5^{\circ}$, $6^{\circ}$, $8^{\circ}$, $10^{\circ}$, and $15^{\circ}$ respectively, are provided.\\
The method also offers good results for the tracking over successive rotations (\textit{i.e.} $90\%$ for $l_\mathrm{ske} \geq 15^{\circ}$), but the performance decreases more rapidly when $l_\mathrm{ske}$ decreases. We note that the best values of the input parameters are smaller than for the first series, because the algorithm tends to reduce the false tracking occurrences by using more restrictive values on larger features matched.\\ 
 
\begin{table}
\caption{Results of the assessments for two successive rotation tracking: the total \textit{positive} rate $R_{p}$ in percent (second line) and the corresponding number of filaments $N_\mathrm{fil}$ (third line), using the best input parameters for different lengths of filament $l_\mathrm{ske}$ (first line).}\label{tbl:results2}
\begin{tabular}{cccccc}     
\hline
\hline
$l_\mathrm{ske}$ & $\geq 5^{\circ}$ & $\geq 6^{\circ}$ & $\geq 8^{\circ}$ & $\geq 10^{\circ}$ & $\geq 15^{\circ}$ \\
\hline
$R_{p}$ & $74\%$ & $77\%$ & $81\%$ & $88\%$ & $90\%$ \\
$N_\mathrm{fil}$ & $1991$ & $1698$ & $1263$ & $924$ & $479$ \\
\hline
\end{tabular}
\end{table} 

\begin{figure} 
 \centerline{\includegraphics[width=1.0\textwidth]{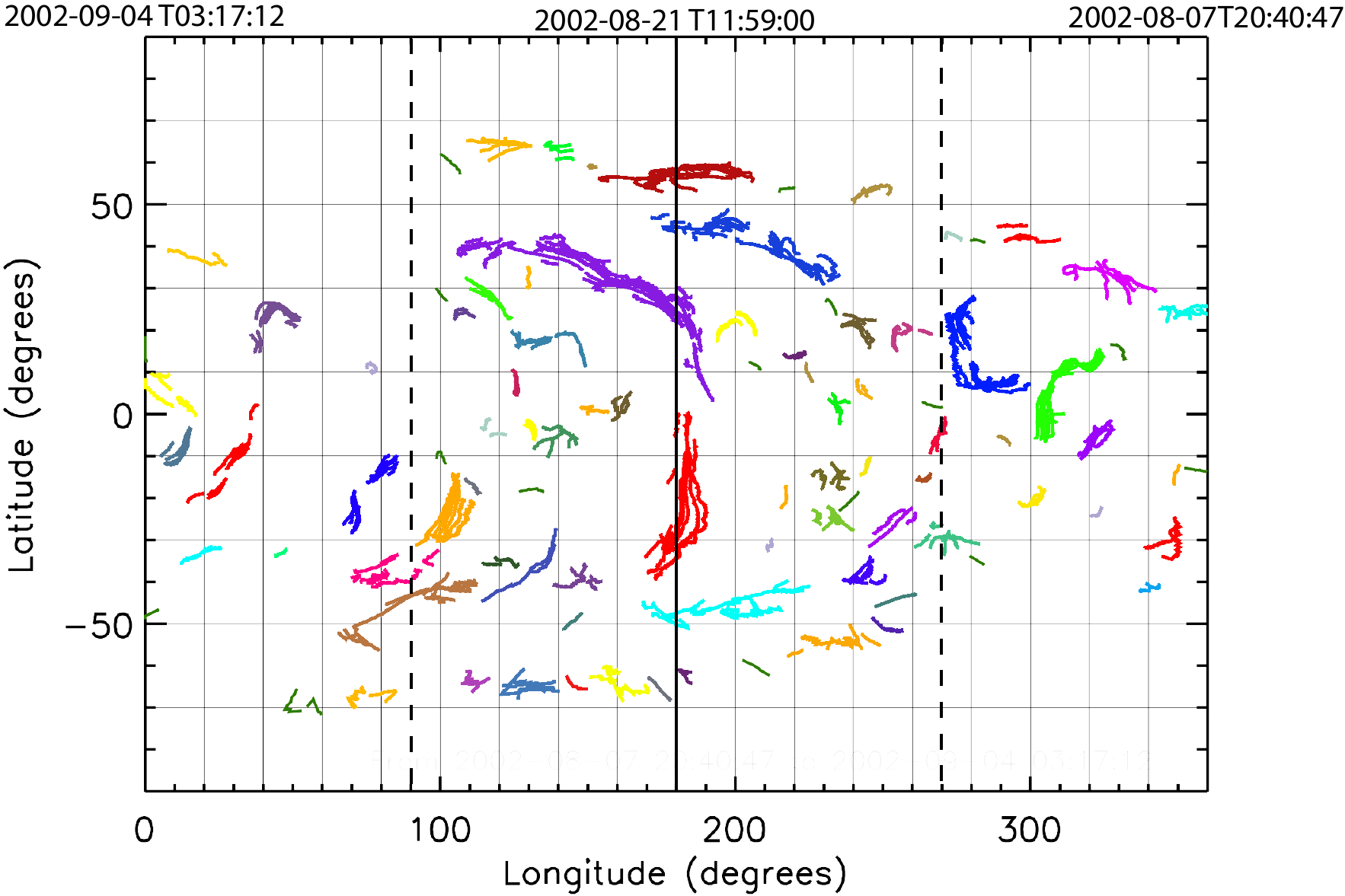}}
  \caption{Same synoptic map as Figure \ref{fig:SynopticMap}, but after the execution of the tracking code using the best input parameters found in the first series. Each group of skeletons that actually correspond to the same co-rotating filament (\textit{i.e.}, which have the same tracking identification number) are plotted with one specific color. Only filaments that have a length greater than $5^{\circ}$ are tracked, and so plotted on the map. The two black dashed lines delimit the longitude range of the image taken at the time $t$, and for which tracking is initially performed.}\label{fig:results}
\end{figure}  

An example of a synoptic map, produced after a run of the tracking code using the best set of inputs parameters found in the first series, is shown on Figure \ref{fig:results}. The colors of skeletons are proportional to the corresponding tracking identification numbers, in such a way that associated features appear with the same color on the plot. In most cases, we can notice that the larger features are well grouped by colors.

\section{Conclusions} 
  \label{S:Conclusions} 

We have presented here a new method to track filaments on solar images. This technique is based on a curve-matching algorithm, applied to the skeletons of filaments in a reference frame rotating with the Sun's surface. The comparison of the automated code results with a representative sample of filaments tracked manually confirms the good performance (close to $\approx 90\%$ ; depending on the length of the skeletons) of the method. Moreover, the results show the good aptitude of the process to identify the main parts of a segmented filament (as a complementary tool to the detection codes, that often offer such a process on each image). Application of the code to track features over two successive rotations gives also good results, but appears to be less efficient, especially for the smaller filaments; improvements to include smaller filaments cases are in progress.\\
For now, a version of the code is already used to provide tracking information to the filaments table of the HFC. Stored content only concerns filaments detected on the Meudon H$\alpha$ spectroheliograms, but efforts are currently made to also include BBSO observations. Joint use of these data sets will allow us to improve the ability of the code to follow filaments, by increasing the temporal resolution (Meudon Observatory provides only one image per day on average), but also will refine the analysis of the filament behaviour, since a low cadence limits the capability of the algorithm to accurately detect the occurrence time of events such as filament disappearances.\\
In addition, several applications of this method are envisaged in the future. Firstly, automated tracking will be a primary step to the creation of filaments synthetic synoptic maps \citep{MouradianASPC1998,AboudarhamASPC2007} ; these maps provides useful information about solar features to the community. Since the detection and tracking codes can be run independently (and insofar as a dedicated detection code can previously provide the pruned skeletons), investigations are in progress to apply this technique to filament extraction codes working on other wavelengths observations \citep{BuchlinSF2A2010}. Finally, an extension to others solar features such as active regions or coronal holes, could be reasonably envisaged using appropriate matching algorithms.

%

%
\begin{acks}
This work has been supported under the Heliospheric Integrated Observatory HELIO project. HELIO is a Research Infrastructures funded under the Capacities Specific Programme within the European Commission's Seventh Framework Programme (FP7; Project No. 238969). The project started on 1 June 2009 and has a duration of 36 months. In addition, the authors want to thank Z.Mouradian from LESIA for the his valuable help, and the referee for helpful comments. The final publication is available at \url{www.springerlink.com}.
\end{acks}

%
%
\bibliographystyle{spr-mp-sola}
\bibliography{Bonnin_SolPhy_2011.bib}  
%
%
%
%

\end{article} 

\end{document}